\documentclass{article}

\usepackage[preprint]{neurips_2023}

\usepackage[utf8]{inputenc} 
\usepackage[T1]{fontenc}    
\usepackage{hyperref}       
\usepackage{url}            
\usepackage{booktabs}       
\usepackage{amsfonts}       
\usepackage{nicefrac}       
\usepackage{microtype}      
\usepackage{xcolor}         
\usepackage{subcaption}
\usepackage[linesnumbered,ruled]{algorithm2e}

\usepackage[utf8]{inputenc} 
\usepackage[T1]{fontenc}    
\usepackage{hyperref}       
\usepackage{url}            
\usepackage{booktabs}       
\usepackage{amsfonts}       
\usepackage{nicefrac}       
\usepackage{microtype}      
\usepackage{xcolor}         

\usepackage{amsmath}
\usepackage{amssymb}
\usepackage{mathtools}
\usepackage{amsthm}

\usepackage{thm-restate}

\usepackage{color-edits}
\addauthor{ev}{blue}

\theoremstyle{plain}

\theoremstyle{definition}

\theoremstyle{remark}

\title{GPT4 is Slightly Helpful for Peer-Review Assistance: A Pilot Study}

\author{%
  Zachary Robertson \\
  Department of Computer Science\\
  Stanford\\
  \texttt{zroberts@stanford.edu} \\
}

\begin{document}

\maketitle

\begin{abstract}
    In this pilot study, we investigate the use of GPT4 to assist in the peer-review process. Our key hypothesis was that GPT-generated reviews could achieve comparable helpfulness to human reviewers. By comparing reviews generated by both human reviewers and GPT models for academic papers submitted to a major machine learning conference, we provide initial evidence that artificial intelligence can contribute effectively to the peer-review process. We also perform robustness experiments with inserted errors to understand which parts of the paper the model tends to focus on. Our findings open new avenues for leveraging machine learning tools to address resource constraints in peer review. The results also shed light on potential enhancements to the review process and lay the groundwork for further research on scaling oversight in a domain where human-feedback is increasingly a scarce resource.
\end{abstract}

\section{Introduction}

The process of peer-review is at the heart of academic research, ensuring the quality and integrity of scholarly publications \citep{burnham1990evolution, spier2002history}. However, as pointed out by \citet{vesper2018peer} and \citet{petrescu2022evolving} the growing volume of research output poses significant challenges to this system, often straining human reviewers' capacities. As a result, there is a compelling need for innovative solutions that can effectively scale with the increasing demand for reviews.

In this vein, the potential of artificial intelligence (AI) and machine learning (ML) tools for augmenting the peer-review process presents an intriguing prospect. More specifically, the Generative Pretrained Transformer (GPT) models developed by OpenAI have demonstrated impressive capabilities in generating human-like text, raising the question of their utility in an academic review setting \citep{brown2020language}.

In this pilot study, we set out to test the hypothesis that GPT-generated reviews could be as helpful as those provided by human reviewers. We employed a controlled experimental design involving 10 participants and compared the helpfulness of human and GPT reviews based on a predetermined scale.

While we acknowledge potential biases in our setup, our aim was to explore the practical viability and potential value of AI-generated reviews, particularly in scenarios where human resources may be constrained. In addition to the human study, we also run quantitative experiments to evaluate the robustness of the model to adversarial attacks in the form of inserted errors.

\begin{figure*}[!t]
    \centering
    ~ 
    \centering
    \includegraphics[width=0.5\linewidth]{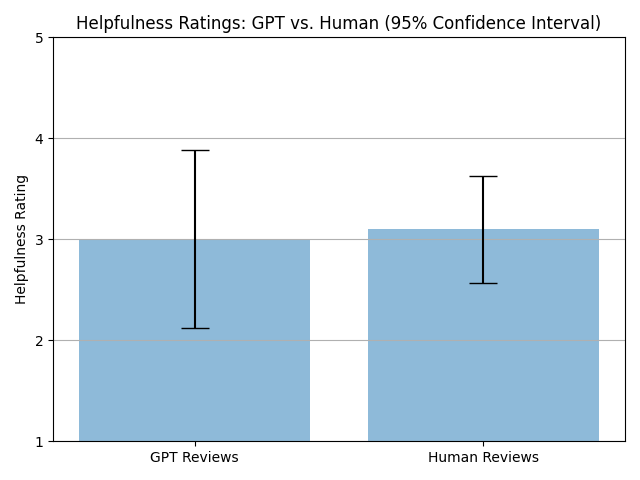}
    
    \caption{Mean Helpfulness Ratings of GPT and Human Reviews. The bar chart illustrates the mean helpfulness ratings for both GPT-generated and human reviews, which both stand at approximately 3 on a scale of 1 to 5. The error bars represent the 95 \% confidence interval, highlighting the variability in ratings for each type of review. Notably, GPT reviews exhibited a larger variance in helpfulness (3 ± 0.96), compared to human reviews (3.1 ± 0.57).}
    \label{fig:results}
\end{figure*}

\section{Methods}

The research methodology adopted for this study involved a mix of structured guidelines for human reviewers, GPT review generation processes, and iterative mechanisms for ensuring appropriate review formatting. Here, we detail the processes involved in carrying out the experiment.

\subsection{Participant Selection and Guidelines}

We selected a total of 10 participants for our study which was conducted via email. Each participant had a paper they authored and wanted feedback for. The participants were paired with each other and instructed to leave a review for an assigned paper following the guidelines and format used by \citet{NeurIPS}. Participants were informed and asked for consent before using GPT to review their papers. One participant declined to have their paper reviewed by GPT.

\subsection{GPT Review Generation}
\label{format}

We use the GPT4 model with a context size of 8k to review papers. Our approach works in the three steps \footnote{In our initial exploration we found only the most powerful language models are capable of producing coherent reviews.}. First, we have GPT generate notes on different sections of the paper that are helpful for generating a review. Second, we have GPT organize these notes into a consistent format. Third, we give GPT formal guidelines for generating a review and then have it generate a review based on the organized notes. We've made the prompts and code publically available \footnote{\url{https://github.com/zrobertson466920/GPT_Auto_Review}}. Also see the appendix.

An iterative process was implemented to ensure proper formatting and coherence of the GPT reviews. We define proper formatting as a review that highlights and discusses the ten points asked in the NeurIPS reviewer guidelines. During the generation process, the following steps were undertaken:

    \begin{enumerate}
        \item Load the paper and generate a new review.
        \item Save the generated review
        \item While review is not in correct format, increment the number of attempts.
    \end{enumerate}

This process was repeated until the generated review was correctly formatted. The number of reattempts for each review was recorded.

\subsection{Review Formatting and Corrections}

We developed a mechanism to ensure proper formatting for both GPT-generated and human reviews. For human reviews, if a review was not properly formatted, the participant was asked to correct it. All reviews, human and GPT-generated, were stored in a worksheet for easy reference and management.

\subsection{Review Assessment}

After all reviews were submitted, each participant received the reviews for their paper and was asked to rate the helpfulness of the review on a scale from 1 to 5, with 1 being "Not at all helpful" and 5 being "Extremely helpful". In the event of a missing review, it was scored as "not at all helpful", considering that the author could not benefit from feedback. The scale given to particpants was as follows:

\begin{enumerate}
\centering
        \item Not at all helpful
        \item Slightly helpful
        \item Moderately helpful
        \item Very helpful
        \item Extremely helpful
\end{enumerate}

This methodological approach, though limited in scope, offered us an initial framework to compare the effectiveness and helpfulness of human and GPT-generated reviews in a controlled setting.

\section{Results}

The outcome of the experiment (Figure \ref{fig:results}) was evaluated based on the helpfulness ratings given to both human and GPT-generated reviews by the participants. We display some sample reviews in the appendix.

\subsection{Statistical Results}

The mean helpfulness rating for both human and GPT-generated reviews stood at 3 on our scale of 1 to 5. In terms of the 95 \% confidence interval, AI reviews exhibited a larger variance in helpfulness, scoring 3 ± 0.96, compared to human reviews, which scored 3.1 ± 0.57. The mean number of reattempts for GPT to output a review in the correct format was 4.8 ± 1.1 with this being defined as discussed in Section \ref{format}. 

\subsection{Observations and Participant Feedback}

We display some sample reviews in the appendix. 

Several participants were able to identify the review generated by GPT. It was noted that while human reviews tended to focus more on specific details within the paper, GPT reviews consistently provided a summary of the paper under review - a feature missing in four out of nine human reviews. Participant feedback also indicated that the general nature of GPT reviews could limit their helpfulness for papers that have already undergone several review rounds. However, for fresh papers, the broad assessment provided by GPT could prove more useful.

\section{Adversarial Robustness Evaluation}

\begin{table*}[!t]
    \centering
    \caption{Transformations. In this table, the first column represents the transformation used, the second column shows an example sentence before the transformation, and the third column displays the sentence after the transformation.}
    \begin{tabular}{|p{4cm}|p{4cm}|p{4cm}|}
        \hline
        \textbf{Transformation} & \textbf{Before} & \textbf{After} \\
        \hline
        Rewrite the following sentence and make one change that makes it unprofessional for a conference paper: & Our goal is to build an explanatory model of an AI system, i.e. a mechanism that translates between its internal representations and/or computations, and some other representational form. & Our objective is to construct a frickin' cool explanatory model for an AI system, something that can convert all its internal models into something more understandable. \\
        \hline
        Rewrite the abstract and negate a key claim: & We develop an instance of this solution for Deep RL agents: Causal Self-Talk. & Our method, dubbed Non-Causal Self-Dialogue (NCSD), encourages the agent to internally echo its actions over time. \\
        \hline
    \end{tabular}
    \label{fig:transform}
\end{table*}

\begin{table*}[!t]
    \centering
    \caption{Robustness Evaluation Results. In this table, the first column represents the model used (GPT4-4k and GPT4-32k), the second column shows the recall results for the abstract swap experiment, and the third column displays the recall results for the informal sentence insertion experiment. We also display 95 \% confidence intervals.}
    \begin{tabular}{|c|c|c|}
        \hline
        \textbf{Model} & \textbf{Abstract Swap} & \textbf{Informal Sentence Insertion} \\
        \hline
        GPT4-4k & $0.7 \pm 0.21$ & $0.6 \pm 0.22$ \\
        \hline
        GPT4-32k & $0.35 \pm 0.21$ & $0.05 \pm 0.1$ \\
        \hline
    \end{tabular}
    \label{fig:robustness}
\end{table*}

We also conducted several follow-up experiments to evaluate the robustness of our auto-peer review model to high-leve and low-level attacks to a paper's contents. The goal was to assess the model's performance and identify limitations or areas for improvement. In this section, we outline the experiments performed and present the findings.

We perform experiments with two models: GTP-4 with a 4k token context-size and GPT-4 with a 32k token context-size. We use 20 papers submitted to NeurIPS for our experiment. We evenly split papers between those accepted and those that were rejected, but opted into public release on OpenReview. The first experiment focused on an abstract change task, where we aimed to rewrite the abstract and negate a key claim. This experiment aimed to test the model's ability to understand the abstract's content as it relates to other sections of the paper. The second experiment focused on the model's attention at the sentence level. We select a sentence at random from each paper and rewrite it to be informal. The purpose was to evaluate whether the model could identify the formality error. 

We display example transformations in \ref{fig:transform} and summarize our findings in Table \ref{fig:robustness}. Overall, we found that GPT-4 was able to identify and raise concerns consistently when the abstract was rewritten, but struggled more with identifying errors inserted at the sentence level. We also find that GPT-4 demonstrated stronger performance when using a smaller context-size. These findings highlight the need for further experimentation and refinement to enhance the model's performance in different scenarios.

\section{Discussion}

Our investigation reveals that AI models, such as GPT, can be effectively utilized to aid the peer-review process, with AI-generated reviews earning moderately favorable feedback from participants. Yet, the restricted scope of our initial study combined with certain limitations restricts us from making absolute conclusions.

A comparative analysis of GPT and human-generated reviews showed that they received comparable average ratings for helpfulness. However, a greater variance in the ratings of the GPT reviews points to possible inconsistency in the performance of the AI model. This discrepancy could be attributed to the need for a few-shot approach in generating reviews. Nevertheless, the AI model's capacity to produce multiple reviews from a consistent source presents a significant advantage over human reviewers, who typically review a limited number of papers for a conference. Therefore, AI-generated reviews could serve as a valuable tool for an AC to balance human-sourced reviews, even if they are not directly used in decision-making processes.

A key constraint of our research stems from its design - our sample size was small as we sought to assess the usefulness of GPT for paper authors. This setup, while realistic in the context of preparing an academic manuscript, might prove challenging to generalize to other stakeholders such as area chairs or the general public. Consequently, our study wasn't designed to evaluate whether GPT would prove useful for peer-review more broadly.

Upon reviewing the AI-generated reviews GPT suggest some further weaknesses. Review patterns show that AI-generated reviews primarily address key contributions, strengths, weaknesses, accuracy, clarity, relationship to prior work, and other additional thoughts. The AI commonly comments on the paper's innovative aspects, practical implications, importance, relevance to the community, and method correctness. There isn't a discernible pattern in the most or least positive reviews, likely due to the limited sample size.

In terms of quality, the reviews are generally of good standard and meet the review prompts, albeit with some inconsistencies and omissions. The AI failed to mention the "Errors" section in some reviews as explicitly instructed in the prompt, focusing more on description and methodology but neglecting to offer strong criticisms or concerns. Several unexpected findings surfaced, such as generating an unclear summaries or not offering valid feedback for certain papers. Furthermore, in certain reviews, the AI posed questions or provided suggestions without clarifying which areas required revision or improvement in clarity. The AI-generated reviews were mostly well-structured, relevant, and helpful in evaluating the scientific papers, even though the AI occasionally missed specific instructions or details from the prompt. There are potential constraints in the AI's capacity to identify and discuss gaps and weaknesses in the submissions.

The AI reviews indicate a commendable performance in analyzing scientific papers but suggest the need for improvement in consistently emphasizing particular aspects like limitations, failures, and negative results. Fine-tuning the AI to better adhere to instructions and maintain a consistent structure and style across reviews could prove beneficial. Future experiments could involve inserting varied or multiple errors in the papers to evaluate the AI's ability to identify several issues concurrently. Additionally, including errors that could significantly alter the interpretation of a paper could provide valuable insight into how well the AI can detect and address them.

\section{Related Work}

While our study presents a novel application of GPT in assisting peer-review, the literature presents various perspectives and approaches to the problem of improving and scaling the review process. We discuss a selection of these works in this section.

\subsection{Human Involvement in Peer-Review}

Despite the advancement in AI and its potential applications, there is a significant amount of research emphasizing the necessity of human involvement in the peer-review process. For instance, a recent study by \citet{schulz2022future} argues that peer-review still requires human judgment, especially when it comes to evaluating the novelty, relevance, and potential impact of research. However, there has been some work on using AI to assist in pre screening papers in the review process to streamline the process \citep{checco2021ai, liu2023reviewergpt}. In particular, \citet{liu2023reviewergpt} is a concurrent work on evaluating GPT for use peer-review, but does not focus on full review generation and does not conduct a human-study. To the best of our knowledge, this work is the first to explore the use of AI to generate peer-review in an end-to-end manner with human evaluation. 

\subsection{Predicting Paper Acceptance or Rejection}

The possibility of predicting paper acceptance or rejection using AI and ML models is an intriguing proposition. Research has shown that AI can be used to predict the outcome of the review process \citep{bharti2021peerassist,bao2021predicting}. While this line of work offers an interesting parallel to our study, ultimately an approach like ours is likely to be more interpretable because the review is left in a structured format with reasoning and verifiable claims.

\subsection{Improving Peer-Review}

The task of augmenting the peer-review process is multifaceted, and an integral part of this involves the strategic application of mechanism design to ensure reviewers and authors faithfully represent the quality of paper submissions. This subject was recently studied by \citet{su2021you}, who introduced the concept of self-reported ranking of papers. Through their rigorous analysis, they established the optimality of this method under certain conditions. This pioneering approach opens new avenues for the exploration of self-regulatory mechanisms within the peer-review process, potentially facilitating more accurate, transparent, and unbiased evaluations.

More broadly, information elicitation without verification stands as a critical issue in broader contexts, not just within the scope of peer review. This problem becomes particularly significant when it comes to evaluating contributions from individuals for which there is no accessible ground truth \citep{miller2005eliciting,waggoner2013information,zhang2014elicitability,kong2019information}.

A noteworthy point of departure in this realm was the peer-prediction method, introduced by \citet{miller2005eliciting}. This technique sought to foster informative feedback in scenarios where objective standards were not readily available. Following this, \citet{waggoner2013information} and \citet{zhang2014elicitability} further expanded on this challenge, highlighting the importance of crafting elicitation mechanisms suited to situations where verification is not achievable. n operate in complex situations where ground truth is unavailable. Most recently, \citet{kong2019information} proposed an information-theoretic framework for designing information elicitation mechanisms that reward truth-telling, providing a novel solution for incentivizing individuals to provide truthful information in the absence of verification.

In sum, our research fits into an ongoing conversation about how to optimize and scale the peer-review process. While AI presents a promising avenue for exploration, these related works remind us of the continuing importance of human involvement, the need for improved mechanisms within the existing process, and the exciting potential for predictive models.

\section{Ethics and Societal Impact}

Given our results, it seems plausible that AI-assisted peer review will surpass the median conference reviewer with further prompt and fine tuning approaches, possibly within a couple of years. This raises several questions we think are worth thinking about. What are the implications of AI-assisted peer review for the academic tradition of volunteer-based reviewing? Most likely, peer-reviewers will begin to rely on automated review assistants as soon as the price becomes economically viable. More research needs to be done to ensure that AI models used in the peer-review process align with our values and objectives for academic research.

More broadly, the automation of academic peer review may affect the role and importance of traditional mentorship in academia. Mentorship traditionally involves imparting knowledge, fostering critical thinking, and nurturing a deep understanding of the subject matter - tasks that require human judgment, empathy, and experience. The automation of certain aspects of this process might lead to concerns about the devaluation of human mentorship or the loss of personalized guidance and support.

While we found GPT's reviews to be helpful in our study, it is crucial to ensure that the AI's judgements are not unfairly biased towards or against certain topics, methodologies, or authors. It is also crucial to ensure accountability in cases where an AI-generated review leads to a wrongful rejection or acceptance of a paper. Finally, the proliferation of AI-assisted reviews could also lead to an over-reliance on technology, which may inadvertently stifle creativity, originality, and diversity in academic research. 

\section{Conclusion}

Our findings, though preliminary, suggest several areas for future research. For instance, it would be valuable to explore whether GPT's performance in providing low-level details of papers could be improved. Further studies involving a larger sample size and fine-tuned models could also help establish more conclusively the potential of AI tools in the peer-review process.

\begin{ack}
I'd like to thank Tatsu Hashimoto for high-level advising of the project and Sanmi Koyejo for assistance finding study participants. I'd also like to acknowledge Rohan Taori and CRFM for providing GPT4 API credits. 
\end{ack}

\newpage

\bibliographystyle{unsrtnat}
\bibliography{references}

\newpage

\appendix

\section{Prompts for Review Generation}

As a reminder the review process occurs in three stages. First, GPT takes notes on sections of the paper. Second, GPT organizes the notes. Third, GPT writes the review.

\textbf{Note gathering instructions:}

\begin{verbatim}
You are reviewing a paper submission for NeurIPS and are in the process of leaving
bullet notes to assist in preparing a peer-review. Key aspects to leave notes on:

0. Look for errors / unclear leaps of logic in the paper's arguments. Typos, syntax
errors, mistakes are all appreciated. Make sure to quote or use concrete references
so comments are easy to find and address in the paper. 
1. Motivations, key contributions, and achievements of the paper.
2. Strengths of the work, such as soundness of claims, significance and novelty of
the contribution, and relevance to the NeurIPS community. Mention relevant prior
works if possible.
3. Limitations of this work along the same axes as above, while being specific
and polite.
4. Commentary on if the claims and method are correct, as well as the empirical
methodology.
5. Commentary on the clarity of exposition of the paper, noting down parts that
need revision to improve clarity.
6. Examination of whether the submission relates the proposed work to prior work
in the literature, and how it differs from previous contributions.
7. If you are at the end of the paper you will see references. Any useful
information you know about these will be helpful to take notes on.
Remember to note you are discussing a reference.

Please provide concise, helpful, and unstructured bullet notes on these points
that will facilitate the process of writing a detailed, constructive, and polite
review later on. These notes should focus on the key points necessary to understand
and evaluate the paper and should pay particular attention to points of concern
that may be unkown to the author or unobvious to someone to who hasn't
read beyond the abstract. Focus on the text, don't copy the abstract.
It is just for context. Text and abstract are below:
\end{verbatim}

\textbf{Note synthesizing instructions:}

\begin{verbatim}
You will collect notes for a single paper submission for NeurIPS and will
summarize the following notes. Notes are for each section of the same paper
and are in a similar format. Key aspects to pay attention to when
synthesizing these notes:

0. Notes may contradict each-other because they were made with a partial
understanding of the paper. Use your best judgement and strive to be as
accurate as possible when synthesizing. 
1. Motivations, key contributions, and achievements of the paper.
2. Strengths of the work, such as soundness of claims, significance and
novelty of the contribution, and relevance to the NeurIPS community.
Mention relevant prior works if possible.
3. Limitations of this work along the same axes as above, while being
specific and polite.
4. Commentary on if the claims and method are correct, as well as the
empirical methodology.
5. Commentary on the clarity of exposition of the paper, noting down
parts that need revision to improve clarity. Emphasize any consistent
errors / unclear leaps of logic. Typos, syntax errors, mistakes are all
appreciated. Be sure to retain quotes / clear references to where error
is in the paper.
6. Examination of whether the submission relates the proposed work to
prior work in the literature, and how it differs from
previous contributions.

Please provide concise, helpful, and unstructured bullet notes on
these points that will facilitate the process of writing a detailed,
constructive, and polite review later on. Focus on things that may
be unkown to the author or unobvious to someone to who hasn't read
beyond the abtract. Veer towards being verbose. Notes and abstract
for this paper are below:
\end{verbatim}

\textbf{Review construction:
}
\begin{verbatim}
You are reviewing a paper submission for NeurIPS. Please make your review as
detailed, specific, and informative as possible; short, superficial reviews that
venture uninformed opinions or guesses are worse than no review since they may
result in the rejection of a high-quality submission. Try to note *any* typos,
confusing passages, technical errors, or examples of incorrect reasoning.
Nitpick results and be skeptical.

Review content is also the primary means by which authors understand their
submissions’ decisions. Reviews for rejected submissions help authors understand
how to improve their work for other conferences or journals. Reviews for
accepted submissions help authors understand how to improve their work for the
camera-ready versions.

The review form will ask you for the following:

1. Summary and contributions: Briefly summarize the paper and its contributions

Summarize the paper motivation, key contributions and achievements in a paragraph.

Although this part of the review may not provide much new information to authors,
it is invaluable to ACs, SACs, and program chairs, and it can help the authors
determine whether there are misunderstandings that need to be addressed
in their author response.

There are many examples of contributions that warrant publication at NeurIPS. 
These contributions may be theoretical, methodological, algorithmic,
empirical, connecting ideas in disparate fields (“bridge papers”), or
providing a critical analysis (e.g., principled justifications of why
the community is going after the wrong outcome or
using the wrong types of approaches.).  

For more examples of what is intended for this question, see section
Examples of Review Content of this guide.

2. Strengths: Describe the strengths of the work. Typical criteria include:
soundness of the claims (theoretical grounding, empirical evaluation),
significance and novelty of the contribution, and relevance
to the NeurIPS community.

List the strengths of the submission. For instance, it could be about the
soundness of the theoretical claim or the soundness of empirical methodology
used to validate an empirical approach. Another important axis is the
significance and the novelty of the contributions relative to what has
been done already in the literature, and here you may want to cite these
relevant prior works. One measure of the significance of a contribution
is (your belief about) the level to which researchers or practitioners will
make use of or be influenced by the proposed ideas. Solid, technical papers
that explore new territory or point out new directions for research are
preferable to papers that advance the state of the art, but only incrementally.
Finally, a possible strength is the relevance of the line
of work for the NeurIPS community.

Clarification of these axes of evaluation, as well as example quotes from
past NeurIPS reviews, can be found in the section Examples of Review Content
of this guide.

3. Weaknesses: Explain the limitations of this work along the same axes as above.

This is like above, but now focussing on the limitations of this work.

Your comments should be detailed, specific, and polite. Please avoid vague,
subjective complaints. Think about the times when you received an unfair,
unjustified, short, or dismissive review. Try not to be that reviewer!
Always be constructive and help the authors understand your viewpoint,
without being dismissive or using inappropriate language. Remember that
you are not reviewing your level of interest in the submission, but its
scientific contribution to the field!

4. Correctness: Are the claims and method correct?
Is the empirical methodology correct?

Explain if there is anything incorrect with the paper. Incorrect claims or
methodology are the primary reason for rejection. Be as detailed, specific
and polite as possible. Thoroughly motivate your criticism so that authors
will understand your point of view and potentially respond to you.

5. Clarity: Is the paper well written?

Rate the clarity of exposition of the paper. Give examples of what
parts of the paper need revision to improve clarity.

6. Relation to prior work: Is it clearly discussed how this work
differs from previous contributions?

Explain whether the submission is written with the due scholarship,
relating the proposed work with the prior work in the literature.
The related work section should not just list prior work, but explain
how the proposed work differs from prior work appeared in the literature.

Note that authors are excused for not knowing about all non-refereed work
(e.g, those appearing on ArXiv). Papers (whether refereed or not) appearing
less than two months before the submission deadline are considered
contemporaneous to NeurIPS submissions; authors are not obligated to make
detailed comparisons to such papers (though, especially for the camera ready
versions of accepted papers, authors are encouraged to).

7. Reproducibility: Are there enough details to reproduce the
major results of this work?

Mark whether the work is reasonably reproducible. If it is not,
lack of reproducibility should be listed among the weaknesses of the submission.

8. Additional feedback, comments, suggestions for improvement and
questions for the authors

Add here any additional comment you might have about the submission,
including questions and suggestions for improvement.

9.  Overall score:

You should NOT assume that you were assigned a representative sample
of submissions, nor should you adjust your scores to match the
overall conference acceptance rates. The “Overall Score” for each
submission should reflect your assessment of the submission’s contributions.

10: Top 5% of accepted NeurIPS papers. Truly groundbreaking work.

9: Top 15% of accepted NeurIPS papers. An excellent submission; a strong accept.

8: Top 50% of accepted NeurIPS papers. A very good submission; a clear accept.

7: A good submission; accept.

I vote for accepting this submission, although I would not be upset if it were rejected.

6: Marginally above the acceptance threshold.

I tend to vote for accepting this submission, but rejecting it would not be that bad.

5: Marginally below the acceptance threshold.

I tend to vote for rejecting this submission, but accepting it would not be that bad.

4: An okay submission, but not good enough; a reject.

I vote for rejecting this submission, although I would not be upset if it were accepted.

3: A clear reject.

I vote and argue for rejecting this submission.

2: I'm surprised this work was submitted to NeurIPS; a strong reject.

1: Trivial or wrong or already known.

10.  Confidence score:

5: You are absolutely certain about your assessment.
You are very familiar with the related work.

4: You are confident in your assessment, but not absolutely certain.
It is unlikely, but not impossible, that you did not understand
some parts of the submission or that you are unfamiliar with some pieces of related work.

3: You are fairly confident in your assessment.  It is possible that
you did not understand some parts of the submission or that you are
unfamiliar with some pieces of related work. Math/other details were not carefully checked.

2: You are willing to defend your assessment, but it is quite likely
that you did not understand central parts of the submission or that you
are unfamiliar with some pieces of related work.
Math/other details were not carefully checked.

1: Your assessment is an educated guess. The submission is not in your area
or the submission was difficult to understand. Math/other details were not carefully checked.

11. Have the authors adequately addressed the broader impact of their work,
including potential negative ethical and societal implications of their work?

Yes, no or only partially. In order to provide a balanced perspective,
authors are required to include a statement of the potential broader impact of their work,
including its ethical aspects and future societal consequences.
Authors should take care to discuss both positive and negative outcomes.
Indicate whether you believe the broader impact section was adequate.

Remember; short, superficial reviews that venture uninformed opinions or
guesses are worse than no review since they may result in the rejection
of a high-quality submission. Verbose is ok.
Emphasize any errors / unclear leaps of logic. Be sure to have quote or
concrete references for errors to help authors address them in the paper:

\end{verbatim}

\section{Sample Reviews}

We show review for three sample papers.

\textbf{Example 1:} A paper titled, "AnoFormer: Time Series Anomaly Detection using Transformer-based GAN with Two-Step Masking" by \citet{shinanoformer}. 

\begin{verbatim}
1. Summary and contributions: This paper proposes AnoFormer, a transformer-based
Generative Adversarial Network (GAN) for unsupervised time series anomaly detection.
It introduces a novel two-step masking strategy, utilizing random masking and
entropy-based re-masking to help the model learn the distribution
of normal time series data effectively. The authors claim that AnoFormer achieves
state-of-the-art results on several time series anomaly detection datasets.

2. Strengths: The proposed method, AnoFormer, is an interesting attempt to adapt
transformer-based GANs for unsupervised time series anomaly detection.
The two-step masking strategy appears to be a novel approach, and the use of
entropy-based re-masking could potentially help identify abnormal points
with high uncertainty.

3. Weaknesses:
- The paper lacks a clear research question or hypothesis.
- The connection between the proposed method and existing literature
is not well established, and it is unclear how this work addresses
gaps in the current research landscape.
- The description of the datasets used for experimentation is
inadequate, and it is uncertain how well they represent
real-world use scenarios.
- The paper does not discuss the computational cost of the proposed
algorithm, nor does it explain the model's scalability with large
datasets or extended time series data.
- The evaluation of the proposed method is limited to a small number
of datasets, raising concerns about generalizability and
applicability to other domains.
- The authors do not provide a clear explanation of why their
proposed model outperforms other models.
- The paper's structure is weak, with different sections
lacking proper organization and coherence.

Correctness: It is difficult to ascertain the correctness of the
claims and methods due to the lack of clarity in the paper.

Clarity: The paper could benefit from better organization and a more coherent
presentation of the proposed method and experimental results.

Relation to prior work: The paper does not clearly discuss how this work differs
from previous contributions, and the connection to existing literature is weak.

Reproducibility: It is difficult to assess the reproducibility of the proposed
method due to the lack of clarity in the paper.

Additional feedback, comments, suggestions for improvement, and questions for the authors:
- Please clarify the research question or hypothesis.
- Strengthen the connection to existing literature and discuss
how this work addresses research gaps.
- Provide more details on the datasets used for experimentation and explain their
relevance to real-world use scenarios.
- Discuss the computational cost and scalability of the proposed algorithm.
- Evaluate the proposed method on a larger number of datasets to demonstrate
generalizability and applicability to various domains.
- Provide a clearer explanation of why the proposed model outperforms other models.
- Improve the structure and organization of the paper for better readability and coherence.

Overall score: 5 (Marginally below the acceptance threshold)

Confidence score: 3 (Fairly confident in the assessment. It is possible that some parts of
the submission were not understood or that the reviewer
is unfamiliar with some pieces of related work.)

Broader impact: Only partially addressed. The authors should include a
more balanced discussion of the potential ethical and societal
implications of their work.
\end{verbatim}

\textbf{Example 2:} A review of a paper titled "Understanding Black-box Predictions via Influence Functions" by \citet{koh2017understanding}.

\begin{verbatim}
1. Summary and Contributions: In this paper, the authors present a novel application
of influence functions, a technique from robust statistics, to understand and explain
the predictions of black-box models. The paper's key contribution is a scalable method
for applying influence functions in modern machine learning (ML) settings. This is 
achieved by developing a simple, efficient implementation that requires only oracle
access to gradients and Hessian-vector products. The authors demonstrate various
applications of influence functions, such as understanding model behavior, debugging
models, detecting dataset errors, crafting training-set attacks, and improving domain
mismatch issues. Primarily, experiments are conducted on linear regression models and
convolutional neural networks (CNNs). The paper is well-structured and relevant to the
areas of explainable AI and model robustness, although it assumes differentiability and
convexity, and the results rely on local changes.

2. Strengths: 
- Novel application of influence functions to modern ML settings
- Clear explanations of methodology and concepts
- Wide range of demonstrated applications
- Relevance to explainable AI and model robustness

3. Weaknesses:
- Assumes differentiability and convexity, which might limit applicability
- Mainly demonstrated on linear regression models and CNNs
- Influence functions can only measure the effects of local changes

4. Correctness: The methodology and claims appear correct, but some claims could
benefit from further empirical evidence.

5. Clarity: The paper is generally clear and well-structured but requires some revisions
to improve readability in certain areas.

6. Relation to prior work: The submission adequately distinguishes itself from previous work
and integrates ideas from influence-based diagnostics, fast cross-validation, and training
point focusing for human attention.

7. Reproducibility: The paper provides enough details for reproducing the major results
and experiments conducted in the work.

8. Additional feedback, comments, suggestions for improvement, and questions for the authors:
- Address the minor typos and confusing passages found in the text
- Expand the application of influence functions to other model types
- Increase empirical evidence, where possible, for supporting claims
- Discuss the limitations of influence functions in more detail, and possibly explore ways
to address them

9. Overall score: 7 (A good submission; accept)
- This submission is a valuable contribution to the field, but it has some limitations that
could be addressed in future works.

10. Confidence score: 4 (Confident in the assessment, but not absolutely certain. It is 
unlikely, but not impossible, that some parts of the submission were not understood or that
some relevant related work is unfamiliar.)

11. Broader Impact: Yes, the authors have adequately addressed the broader impact of their
work, discussing both positive and negative implications of their influence function approach
in machine learning model analysis and interpretation.
\end{verbatim}

\textbf{Example 3:} A review of a paper titled "CHIMLE: Conditional Hierarchical IMLE for Multimodal Conditional Image Synthesis" by \citet{peng2022chimle}.

\begin{verbatim}
1. Errors: No errors were found.

2. Motivations, key contributions, and achievements of the paper:
The paper proposes a new method called Conditional Hierarchical IMLE (CHIMLE)
for multimodal conditional image synthesis. The method improves upon the prior
best IMLE-based approach, cIMLE, by achieving high sample efficiency and quality.
The main proposed ideas include partitioning of the latent code, partial evaluation
of latent code components, and iterative construction of latent code.
CHIMLE outperforms cIMLE, as well as GAN-based and diffusion-based methods,
across four tasks: night-to-day, $16\times$ single image super-resolution,
image colorization, and image decompression.

3. Strengths of the work: The method successfully addresses the tradeoff
between sample quality and sample efficiency in IMLE, achieving state-of-the-art
results in various tasks. The proposed techniques (partitioning of the latent code,
partial evaluation of latent code components, and iterative construction of latent code)
are novel and effective.

4. Limitations: CHIMLE can only generate output images that are
similar to the training data. For example, if there were no green apple
in the training data, the model would not be able to generate green apples.

5. Correctness of claims and methods: The claims and methods presented
in the paper appear to be correct.

6. Clarity of exposition: The paper is well-written and clear. It presents
the ideas and methodology in a coherent and organized manner.

7. Relation to the prior work: The paper relates the proposed CHIMLE method
to previous works focusing on IMLE-based multimodal image synthesis (cIMLE),
as well as GAN-based and diffusion-based methods. It also compares the
performance of CHIMLE to that of leading general-purpose and task-specific
methods in the literature.

8. Additional thoughts: The results indicate that CHIMLE has the potential
to improve upon existing methods for multimodal conditional image synthesis.
Moreover, the modular design of the Tower Implicit Model (TIM) architecture
allows for easy expansion, though increased model capacity could have
environmental implications. Future work could explore further improving
sample diversity and addressing limitations with generating images
dissimilar to the training data.
\end{verbatim}

\end{document}